\documentclass[journal,table]{IEEEtran}
\IEEEoverridecommandlockouts
\usepackage{cite}
\usepackage{amsmath,amssymb,amsfonts}
\usepackage{graphicx}
\usepackage{textcomp}
\usepackage[xcdraw]{xcolor}
\usepackage{cite}
\usepackage{multirow}
\usepackage{arydshln}
\usepackage{graphicx}
\usepackage{svg}
\usepackage{algorithm}
\usepackage{algpseudocode}
\usepackage{gensymb}

\usepackage{amsmath}
\usepackage{booktabs,tabulary}%table parameter
\def\BibTeX{{\rm B\kern-.05em{\sc i\kern-.025em b}\kern-.08em
    T\kern-.1667em\lower.7ex\hbox{E}\kern-.125emX}}
\begin{document}

% \title{Power Estimation using Physics-Inspired DNN's in Incompletely Observed Power Systems }
\title{Data-Driven Flow and Injection Estimation in PMU-Unobservable Transmission Systems}

% \author{\IEEEauthorblockN{1\textsuperscript{st} Satyaprajna Sahoo}
% \IEEEauthorblockA{\textit{School of Electrical, Computer and Energy Engineering} \\
% \textit{Arizona State University}\\
% Tempe, United States of America \\
% \and
% \IEEEauthorblockN{2\textsuperscript{nd} Anwarul Islam Sifat}
% \IEEEauthorblockA{\textit{School of Electrical, Computer and Energy Engineering} \\
% \textit{Arizona State University}\\
% Tempe, United States of America \\
% \and
% \IEEEauthorblockN{3\textsuperscript{rd} Anamitra Pal}
% \IEEEauthorblockA{\textit{School of Electrical, Computer and Energy Engineering} \\
% \textit{Arizona State University}\\
% Tempe, United States of America \\

% \author{\IEEEauthorblockN{Satyaprajna Sahoo\IEEEauthorrefmark{1}, \textit{Student Member, IEEE}, Anwarul Islam Sifat\IEEEauthorrefmark{2}, \textit{Graduate Member, IEEE}, Anamitra Pal\IEEEauthorrefmark{3}, \textit{Senior Member, IEEE}}

\author{\IEEEauthorblockN{Satyaprajna Sahoo, \textit{Student Member, IEEE}, Anwarul Islam Sifat, 
\textit{Member, IEEE}, and Anamitra Pal, \textit{Senior Member, IEEE}}

\thanks{This work was supported in part by  the National Science Foundation (NSF) grant under Award ECCS-2132904.
The authors are with the School of Electrical, Computer, and Energy Engineering (ECEE) at Arizona State University (ASU).}% <-this % stops a space
% \thanks{}% <-this % stops a space
% \thanks{Manuscript received April 19, 2005; revised August 26, 2015.}}

\vspace{-2em}
}

% \markboth{Journal of \LaTeX\ Class Files,~Vol.~14, No.~8, August~2015}%
% {Shell \MakeLowercase{\textit{et al.}}: Bare Demo of IEEEtran.cls for IEEE Journals}

\maketitle

\begin{abstract}
% With the continued proliferation of renewable generation as well as the increasing frequency and intensity of extreme weather events, a need has been felt for fast and accurate power flow and power injection estimation in the power system. 
% Time-stamped measurements obtained from phasor measurement units (PMUs) can satisfy this need provided the challenge of real-time unobservability can be tackled. 
Fast and accurate knowledge of power flows and power injections is needed for a variety of applications in the electric grid.
Phasor measurement units (PMUs) can be used to directly compute them at high speeds; however, a large number of PMUs will be needed for computing \textit{all} the flows and injections.
Similarly, if they are calculated from the outputs of a linear state estimator, then their accuracy will deteriorate due to the quadratic relationship between voltage and power.
% However, if power flows and injections are calculated from the outputs of a state estimator, then their accuracy will deteriorate due to the quadratic relationship between voltage and power.
% Although phasor measurement units (PMUs) can be used to directly compute the flows and injections at high speeds, a large number of PMUs will be needed for computing \textit{all} the flows and injections.
% One can compute power flows and injections at high speeds (e.g., 30 samples/second) directly from phasor measurement unit (PMU) data.
% However, to estimate \textit{all} the power flows/injections, a large number of PMUs will be needed.
This paper employs machine learning to perform fast and accurate flow and injection estimation 
% at PMU timescales 
in power systems that are sparsely observed by PMUs.
% We first demonstrate that state estimation-based approaches for flow and injection estimation suffer due to the quadratic relationship between voltage and power. Then, 
We train a deep neural network (DNN) to learn the mapping function between PMU measurements and power flows/injections. 
% To 
% further 
% improve performance, we 
% We also 
% embed the physics of the system 
The relation between power flows and injections is incorporated
into the DNN by adding a linear constraint to its loss function.
% Then, to improve performance, we embed the physics of the system by ...
The results obtained using the IEEE 118-bus system indicate that the proposed approach performs more accurate flow/injection estimation in \textit{severely} unobservable power systems compared to other data-driven methods. 
% Studies conducted on additional PMU placement show that PIC-DNN can increase  estimation accuracy by up to 14\%.
% Traditional real-time power system data acquisition via supervisory control and data acquisition (SCADA) are not time synchronized. Therefore, system wide analysis using SCADA is difficult.
% Time synchronized data in power systems is provided by phasor measurement units (PMU), however in reality PMU's are not located at each bus in the network. This creates an unobservability issue which cannot be solved by standard weighted least squares (WLS) state estimation approaches.
% Moreover, LSE and other state estimation approaches focus on voltage phasors, which while can be used to calculate other state parameters like power injection, generally result in bad estimates, due to errors in individual states combined.
% This paper proposes a deep learning based method to simultaneously estimate the power flows and injections in a network. 

\end{abstract}


\IEEEpeerreviewmaketitle

\vspace{-1pt}
\section{Introduction}

Knowledge of active power flows and power injections is fundamental for the reliable, resilient, and economic operation of the electric grid. Traditionally, their knowledge had been used to determine the cost for buying/selling energy as well as for performing vulnerability/security assessment \cite{chen2020tracing,biswas2020graph}. More recently, with the growing frequency and intensity of extreme weather events as well as increasing penetration of renewable energy resources, a need has been felt for \textit{high-speed tracking} of the power flowing through critical equipment, such as transformers, to determine their health \cite{azmi2017evolution}.

Power flows and injections can be \textit{directly} computed at high speeds from the outputs of a phasor measurement unit (PMU). However, a large number of PMUs will be needed to estimate \textit{all} of them. 
They can be calculated \textit{indirectly} from the outputs of a linear state estimator.
However, the quadratic relationship between voltage and power deteriorates the quality of flow/injection estimates considerably (see Table \ref{table1} in Section \ref{section2} for a sample illustration).
% Power flows and injections can be calculated from the outputs of a state estimator (i.e., in an \textit{indirect} manner). However, the quadratic relationship between voltage and power deteriorates the quality of flow/injection estimates considerably (see Table \ref{table1} in Section \ref{section2} for a sample illustration).
% The power flows/injections can be \textit{directly} computed at high-speeds from the outputs of a phasor measurement unit (PMU). However, a large number of PMUs will be needed to estimate \textit{all} the flows and injections. 
% \cite{pagnier2021physics}. 
This paper presents a physics-inspired machine learning (ML) formulation to perform data-driven power flow and injection estimation in PMU-unobservable power systems. We focus on the transmission system since most of them already have some PMUs installed. However, the proposed methodology is generic enough that it can be applied to distribution systems as well.

\textit{Literature Review:} 
% The state-of-the-art when it comes to high-speed (sub-second) power flow and/or power injection estimation is fairly limited. 
In \cite{amini2014determination}, a minimum
variance unbiased estimator was developed to calculate the power flows using the state estimation approach. However, the analysis was restricted to DC power flows.
A linearly constrained least-squares optimization problem was formulated in \cite{al2016estimating} to estimate nodal power injections that did not violate power flow constraints. However, no information was provided regarding how the formulation could perform high speed estimation in unobservable power systems.
A methodology to learn the topology and estimate the injection statistics in distribution systems with unobservable nodes at PMU timescales was developed in \cite{deka2020joint}. However, it required the unobservable nodes to be non-adjacent, which may not always be the case.   

\textit{Main Contributions:} This paper proposes the use of ML to simultaneously estimate all power flows and injections in a system that is sparsely observed by PMUs. The ML model is built using deep neural networks (DNNs).
The flows and injections in a power system are related by the law of conservation of energy. This knowledge is embedded into the DNNs as a \textit{linear constraint}; the resulting ML model is referred to as a physics-inspired constrained-DNN (PIC-DNN).
The offline training is performed using slow timescale historical data obtained from the supervisory control and data acquisition (SCADA) system. The online implementation only uses PMU data, ensuring high speed of estimation.
The superior performance of the proposed model over classical as well as other ML models is demonstrated for different numbers of PMUs placed in the IEEE 118-bus system.

\section{Sources of Error in Power calculated from State estimates} \label{section2}
% \section{Error Propagation during Power Flow and Injection Estimation from State Estimation}\label{section2}
% \begin{itemize}
%     \item Error analysis and propagation in LSE
%     \item why LSE is not good enough
%     \item why ML 
% \end{itemize}

The most traditional way of determining electrical quantities in a power system is through 
% the process of 
state estimation. 
The widely used static state estimator provides an estimate of the voltage phasor (magnitude and angle) of every bus of the system.
When a system is completely observed by PMUs, then by only employing PMU data one can perform \textit{linear} state estimation (LSE); LSE is faster and more accurate than SCADA-based state estimation \cite{chatterjee2015partitioned}.
However, there are two issues with computing power flows/injections from the outputs of LSE: (a) quadratic relationship between voltage and power, and (b) non-Gaussian noise in PMU measurements. These two issues are elaborated below.

\textit{Quadratic Relationship:} Power flowing through a branch located between bus $i$ and bus $j$, denoted by $p_{ij}$, is related to the voltages of the two buses by the equation shown below:
\begin{equation} \label{eq1}
p_{ij} = g_{ij}({v_i}^2 - v_i v_j \cos{\delta_{ij}}) - b_{ij}(v_i v_j \sin{\delta_{ij}})
\end{equation}

In \eqref{eq1}, $v_i$ and $v_j$ denote the voltage magnitudes of buses $i$ and $j$, $\delta_{ij}$ denotes the voltage angle difference between the two buses, and $g_{ij}$ and $b_{ij}$ are the real and imaginary components of the admittance matrix, respectively.
Similarly, the power injection at bus $i$, denoted by $p_{i}$, can be expressed as,
% \begin{equation}\label{eq2}
% p_{i} = \sum_{k=1}^{n} v_i v_k (g_{ik} \cos{\delta_{ik}} + b_{ik} \sin{\delta_{ik}})
% \end{equation}
% where $n$ denotes the total number of buses in the system. 
\begin{equation}\label{eq2}
p_{i} = \sum_{j=1}^{br_{i}} p_{ij} = \sum_{k=1}^{bu} v_i v_k (g_{ik} \cos{\delta_{ik}} + b_{ik} \sin{\delta_{ik}})
\end{equation}
where $br_{i}$ denotes the total number of branches incident on bus $i$, and $bu$ denotes the total number of buses in the system.
It is clear from \eqref{eq1} and \eqref{eq2} that power is proportional to the square of the voltage.
Therefore, even a small error in the voltage estimates will get amplified when used for calculating the power flows and power injections.
% Furthermore, since power injection at a bus is a function of the number of branches incident on that bus, estimating the injections 
% % (especially if they are small) 
% from the voltages can result in numerical problems.

% (explanation) Intuitively, a highly connected bus that is unobserved, but is adjacent to multiple PMU buses is expected to have good estimation accuracy. But \eqref{eq2} implies the contrary. The measurement errors in PMU buses add up to increase the estimation error of a well connected unobservable bus. This is not the case with  data driven models, which would decrease the error with the availability of more data.

\textit{Non-Gaussian Noise:} LSE is most commonly performed using the \textit{least squares} method. However, least squares is the solution to the \textit{maximum likelihood estimation} problem under Gaussian noise environments \cite{varghese2022transmission}. Recently, it has been demonstrated that PMU data has non-Gaussian noise \cite{wang2018assessing}. This, in combination with the previously-mentioned source of error, implies that an \textit{indirect} estimation of power flows/injections (by finding the voltages first using state estimation) can result in lower accuracy. 
This implication is further reinforced through a numerical analysis performed on the IEEE 118-bus system. The results are provided in Table \ref{table1} below.

Table \ref{table1} shows the estimation performance (in terms of root mean square error) when the output of a linear state estimator is used to calculate the power flows and injections in the presence of no noise, 1\% total vector error (TVE) Gaussian noise, and 1\% TVE non-Gaussian noise. The non-Gaussian noise is described by a two-component Gaussian mixture model (GMM); its parameters are given in the last paragraph of Section \ref{CS-1}. 
It is observed from the table that the estimation error increased by at least $60\%$ when the distribution of the noise changed from Gaussian to non-Gaussian.
% , despite the magnitude of noise remaining the same. 
% This reduces the reliability of LSE based estimates in real time implementations. 
% It is clear from the table that due to the two sources of error identified above, despite a nominal error in voltage magnitude and angle estimates, there is a considerable deterioration in the power flow and injection estimates. 
Furthermore, a purely PMU-based LSE requires complete observability of the system by PMUs (for obtaining the results shown in Table \ref{table1}, PMUs were placed at 32 optimal locations that completely observed the system \cite{pal2013pmu}).
This requirement of complete observability is another concern of the LSE-based approach.

\begin{table}[ht]
\caption{LSE results for voltage and power estimates in IEEE 118-bus system}
\label{table1}
\centering
\begin{tabular}{|
>{\columncolor[HTML]{EFEFEF}}c |c|c|c|}
\hline
{\textbf{}}                                                      & \cellcolor[HTML]{EFEFEF}{\textbf{No noise}} & \cellcolor[HTML]{EFEFEF}{\textbf{\begin{tabular}[c]{@{}c@{}}Gaussian \\ noise (1\% TVE)\end{tabular}}} & \cellcolor[HTML]{EFEFEF}{\textbf{\begin{tabular}[c]{@{}c@{}}Non-Gaussian \\ noise (1\% TVE)\end{tabular}}} \\ \hline
\textbf{\begin{tabular}[c]{@{}c@{}}Voltage \\ magnitude (pu)\end{tabular}}  & 4.36 e-13                                          & 0.0036                                                                                                    & 0.0060                                                                                                        \\ \hline
\textbf{\begin{tabular}[c]{@{}c@{}}Voltage angle \\ (degrees)\end{tabular}} & \cellcolor[HTML]{FFFFFF}1.88 e-11                  & \cellcolor[HTML]{FFFFFF} 0.1764                                                                            & \cellcolor[HTML]{FFFFFF}0.2890                                                                                \\ \hline
\textbf{\begin{tabular}[c]{@{}c@{}}Power \\ flow (MW)\end{tabular}}    & 1.46 e-10                                          & 1.3382                                                                                                    & 2.2273                                                                                                       \\ \hline
\textbf{\begin{tabular}[c]{@{}c@{}}Power \\ injection (MW)\end{tabular}}         & 2.69 e-10                                          & 1.3613                                                                                                  & 2.2912                                                                                                        \\ \hline
\end{tabular}

\end{table}

Note that a PMU measures voltage and current phasors at the location where it is placed (subject to its measurement channel limitations \cite{pal2017PMU}). This means that the outputs of a PMU can be used to \textit{directly} estimate all the power flows (and subsequently the power injections using \eqref{eq2}). However, this translates to the well-known \textit{minimum vertex cover} problem \cite{anderson2012minimum}, which requires placing even more PMUs than those required for complete observability for LSE (which is the solution to the \textit{minimum dominating set} problem). For example, PMUs must be placed at 61 locations in the IEEE 118-bus system to estimate all the power flows using only PMU data.
In this paper, we employ ML to directly estimate all the power flows and injections at PMU timescales and with reasonable accuracy while placing 
% a much less number of PMUs.
significantly fewer PMUs.

\section{Physics-Inspired ML for Power estimation in PMU-unobservable Systems}\label{section3}
Data-driven approaches have been shown to attain considerable success for a variety of power system estimation problems \cite{khodayar2021deep}.
Some popular data-driven approaches include linear regression (LR), support vector regression (SVR), and DNN-based regression. 
As the linear relation between the input (PMU measurements) and output (power flows and power injections) for transmission systems that are sparsely observed by PMUs is not guaranteed, LR may not be a good fit for the problem considered in this paper.
Similarly, the time complexity of SVR is quadratic w.r.t. the number of training samples \cite{abdiansah2015time}, which makes it challenging to implement it in a large network with wide variations in features.
Therefore, we employ DNNs to perform direct estimation of power flows and injections from PMU measurements.

In Section \ref{DNN_Unobservability}, we show how DNN-based regression can be used to overcome the unobservability problem associated with purely PMU-based estimation. In Section \ref{PIC-DNN}, we embed the physical law that relates flows and injections into the DNN framework to create the proposed PIC-DNN. 
% Data driven approaches have seen a wide variety of success in power system state estimation problems \cite{dehghanpour2018survey}. Our requirement for high speed real time estimation under minimum input data stipulates the application of deep non-linear models. In the case of linear regression, the linear correlation between input and output data is not guaranteed. Furthermore, the time complexity of support vector regression (SVR) has a quadratic growth proportional to the number of training samples \cite{abdiansah2015time}, which makes it challenging to implement in a large network with wide variations in features. Therefore, in this section, we will focus on physics-inspired deep neural networks (DNN) to improve the performance of direct estimation of power flows and injections.

\vspace{-1em}
\subsection{DNN Regression to Overcome Unobservability}
\label{DNN_Unobservability}
Recently, we have demonstrated the ability of DNNs to perform high-speed time-synchronized static state estimation in distribution systems that are sparingly observed by micro-PMUs \cite{azimian2021time,azimian2022state}.
We did this by using the slow timescale historical smart meter data to create a mapping function between the fast timescale micro-PMU measurements and the states.
The mapping function was learned using a DNN because it has excellent approximation capabilities.
Here, we leverage this concept to perform power flow and injection estimation from sparsely-placed PMUs in transmission systems.

% We start by making three assumptions: (1) the system is completely observed by SCADA, (2) we have access to historical SCADA data and system information for a sufficient time-period (say, few months), and (3) PMUs are already placed on select buses of the system (say, highest voltage buses). 
We start by making two assumptions: (1) we have access to historical SCADA data and system information (e.g., topology) for a sufficient time-period (say, a few months), and (2) PMUs are already placed on select buses of the system (say, highest voltage buses).
Using historical SCADA data and system information, we solve the power flow problem to produce voltage and current phasor measurements corresponding to the locations where PMUs are placed.
The power flow problem also generates the flow and injection information that matches the PMU measurements.
% This ensures that the created PMU measurements match the flow and injection information used to solve the power flow problem. 
Then, we train a DNN whose inputs are the PMU measurements and outputs are the flows and injections.
% injections, both of which are obtained from the solution to the power flow problem.
Finally, during online implementation, streaming data from a select few PMUs is fed into the trained DNN to estimate all the flows and injections at PMU timescales.
% Finally, streaming PMU data is fed into the trained DNN during online implementation to estimate the power flows and injections at PMU timescales using measurements coming from those select few PMUs.
In this way, a DNN can perform purely PMU-based estimation without needing PMUs to completely observe the system.
This DNN model is henceforth referred to as a Direct DNN.

% Deep neural network based formulations for overcoming unobservability in distribution systems is covered in \cite{azimian2021time}. Their formulation retained the high speed PMU measurements to produce time synchronised estimates for power system states with very few PMUs. This paper leverages their concept for power flow and injection estimation in transmission systems. Our proposed methodology is covered in the latter sections.

\vspace{-1em}
\subsection{Incorporating Physics-based Constraints into DNN}
\label{PIC-DNN}
The Direct DNN model developed in Section \ref{DNN_Unobservability} can estimate the power flows and power injections \textit{independently} from the phasor measurements coming from a select few PMUs.
However, as seen in \eqref{eq2}, the flows and injections are related by the law of conservation of energy. To account for this physical law, we modify the DNN architecture by appending a linear constraint to its loss function; the resulting model, called PIC-DNN, is described below.
In the following, small-case letters indicate vectors, upper-case letters indicate matrices, and the symbol $\hat{}$ indicates estimates.

% The objective (loss) function of a regular DNN with weights, $W$, and biases, $b$, is defined as,
% \begin{equation}\label{vanillaopti}
%      \underset{W,b}{\min} (y - \hat{y})^2 ,\;\;  \; \hat{y} = F(z, W, b)
% \end{equation}
% where, $F(z,W,b)$ denotes the function that maps the inputs, $z$, to the output, $y$. For PMU-based power flow and injection estimation, $z=[V \; ; \; I]_{n \times 1}$ and $y=[P_f \; ; \; P_{in}]_{m \times 1}$.
For an input, $z$, and output, $y$, a regular DNN with weights, $W$, and biases, $b$, tries to find a function, $F$, that minimizes the difference between $y$ and $\hat{y}$, where $\hat{y} = F(z, W, b)$. 
The Direct DNN model described in Section \ref{DNN_Unobservability} performs this minimization using a mean squared error (MSE) loss function, where $z=[V \; ; \; I]_{n \times 1}$ and $y=[P_f \; ; \; P_{in}]_{m \times 1}$.
% Note that for the problem being solved here, $z=[V \; ; \; I]_{n \times 1}$ and $y=[P_f \; ; \; P_{in}]_{m \times 1}$. The Direct DNN model performs this minimization using a mean squared error loss function.
To incorporate the law of conservation of energy into the Direct DNN model, we start by defining the following loss function:
% The law of conservation of energy is incorporated into this loss function by modifying it as shown below:
\begin{equation} \label{DNN_opti-linear}
\begin{aligned}
\min_{W,b} \quad & (y - \hat{y})^2 \\
\textrm{s.t.} \quad & \hat{P}_{in} = A \times \hat{P}_f   \\
\textrm{where} \quad & A =  \begin{bmatrix} a_{11} & a_{12} & \hdots & a_{1m_b}  \\
            a_{21} & a_{22} & \hdots & a_{2m_b}\\
            \vdots &  & \ddots & \vdots\\
            a_{(m-m_b)1} & \hdots &  & a_{(m-m_b) m_b}
            \end{bmatrix} 
\end{aligned}
\end{equation}

In \eqref{DNN_opti-linear}, $A$ is a sparse matrix ($a_{ij} \in \{0,1\}^{(m-m_b) \times m_b}$) which denotes the power flows that must be added to calculate the power injection of a bus, and $m_b$ and $m - m_b$ denote the number of branch power flows and bus power injections in the system, respectively.
% bus. The number of branch power flows and bus power injections 
% % (or current measurements) 
% in the system are denoted by $m_b$ and $m - m_b$,  respectively. 
Now, power injections can be removed from the output features of the DNN and calculated from the power flow estimates in the following way:
\begin{equation} \label{PowFlow2PowInj}
    \hat{y} = B \times \hat{P}_f, \; \; \; B = [\mathbb{I}_{m_b}  \; ; \; A] \in \{0,1\}^{m \times m_b}
\end{equation}

The conversion matrix, $B$, is the vertical concatenation of an appropriately sized identity matrix with $A$, and enables us to get both flow and injection estimates. 
Thus, by modifying the loss function of the conventional DNN in the manner shown in \eqref{DNN_opti-linear} and \eqref{PowFlow2PowInj}, we are able to simultaneously minimize the error in the estimates of power flows and power injections while also accounting for the law that relates the two.  

There are several ways to implement the above-mentioned constrained formulation inside a DNN model. A simple strategy would be to add a convex optimization layer to the model output. However, the inclusion of such a layer will result in solving two different optimization problems simultaneously during training, which can become computationally burdensome.
A more intuitive strategy is to add a static layer after the DNN output layer that calculates the combined $\hat{y}$ from the estimated $\hat{P}_f$ and backpropagates the net error. The updated loss function is given by,

% \begin{equation}
% \label{eq6}
% \begin{aligned}
% % G(z) = 
% \min_{W,b} \quad & ( (B^TB )^{-1}B^T \times y - \hat{y})^2
% \end{aligned}
% \end{equation}

\begin{equation}
\label{eq6}
\begin{aligned}
\min_{W,b} \quad & (B^TB )^{-1}B^T ({y} - B \times \hat{P}_f)^2\\
\textrm{where} \quad & \hat{P}_f = F(z,W,b)
\end{aligned}
\end{equation}

The static weight, $(B^TB )^{-1}B^T$, is the Moore–Penrose inverse of the conversion matrix, $B$. Using \eqref{eq6}, we are not only able to reduce the number of DNN output variables to just the power flows but also implicitly include the linear constraints into the loss function without needing a separate convex optimization layer. 

Next, the training dataset, $y_{train}$, which is only composed of the branch power flows, is split into multiple bins in accordance with the variations that occur in the flows.
This improves the accuracy of the DNN because it now has to estimate features that have similar variations.
% % To improve the estimation accuracy further, we bin the branch power flows according to their variations in the training database. 
% To improve the estimation accuracy further, the training dataset, $y_{train}$, which is now only composed of the branch power flows, is split into multiple bins in accordance with the variations that occur in the flows.
% % are binned by their standard deviations 
% This is done because features that have similar variations can be estimated with a higher accuracy by a single DNN model. 
The selection of the optimal number of bins, $n_{bin}$, involves a trade-off. Increasing the number of bins homogenizes the output data, 
% in standard deviation, and hence improves 
thereby improving the accuracy of the model.
However, unrestricted increment in bin count can increase computational burden while decreasing performance utility.
% Empirical tests indicate $n_{bins}  =  5$ to be an appropriate value for power flow estimation.

In summary, the two ways in which we modify the Direct DNN model to create the PIC-DNN model are: (1) modifying the loss function to include a linear constraint, and (2) binning $y_{train}$ in accordance with the observed variations in the power flows in the training database.
The overall implementation of PIC-DNN is shown in \textbf{Algorithm \ref{alg:cap}}.
The results obtained when the PIC-DNN as well as other ML models are used to estimate power flows and injections in the IEEE 118-bus system is provided in the next section.

\begin{algorithm}
\caption{Implementation of PIC-DNN}\label{alg:cap}
\textbf{Input:} Noisy measurements, $z$ \\
\textbf{Output:} Power flow and injection estimates, $\hat{y}$

\begin{algorithmic}[1]
\Procedure{}{}  
\State Define $A$, $B$, $n_{bin}$, $epoch$ 
% \State $y_{train} \gets (B^TB)^{-1}B^T \times y$ 
\State Split $y_{train}$ into $n_{bin}$ bins 

\For{i=1 \textbf{to} $epoch$}{}{}
    \For{j=1 \textbf{to} $n_{bin}$}{ }{}
        \State $\hat{p}_j \gets F_j(z,W_j,b_j)$
    
    \EndFor
    \State $\hat{P_f} \gets [\hat{p}_1, \hat{p}_2, ... , \hat{p}_{n_{bin}}] $ 
    \State $\hat{y} \gets B \times \hat{P}_f$
    % \State $\epsilon \gets (y - \hat{y})^2$
    \State $\delta  \gets   (B^TB)^{-1}B^T (y - \hat{y})^2$
    \State \textbf{backpropagate} $\delta$
    
\EndFor
\EndProcedure
\end{algorithmic}
\end{algorithm}

\section{Results}\label{section4}

% Consequently, two case studies are performed, and their results shown. The first case-study focuses on comparing the performance and reliability of the proposed model against other machine learning and conventional estimation models.

% The second case-study involves an in depth analysis of optimal PMU placement of all models considered in the first case-study, and compares the improvement in performance when PMUs are added in these optimal locations. 

The proposed PIC-DNN model for power flow and injection estimation was applied to the IEEE 118-bus system. This system has 99 loads, 54 generators, 11 high voltage (HV) buses, and 186 branches.
PMUs were assumed to be placed by default on the 11 HV buses such that all the branches coming out of these buses was directly monitored by them.
% The proposed power flow and injection estimation model are extensively tested on data from the 118 bus system. The configuration of the 118 bus system, (a) 118 buses, 11 of which are high voltage buses, (b) 186 branches consisting of 7 transformers, 99 loads, and 54 generators. 
To obtain branch power flows and bus power injections for this system, we solved an AC optimal power flow (ACOPF) using MATPOWER \cite{zimmerman2010matpower}. The process is similar to the approach in \cite{azimian2022state}, where a distribution kernel was fit over historical slow timescale data, and then multiple samples were drawn from it to generate realistic load variation data.
For the given application, the source of slow timescale data was the SCADA system. However, SCADA data is not available for the IEEE 118-bus system. Therefore, we superimposed the variations of similar loads found in the publicly available 2000-bus Synthetic Texas system \cite{birchfield2017grid} onto loads of the IEEE 118-bus system.
Doing so ensured that our load variations were realistic.
Afterward, the outputs of the ACOPF were used to train the ML models.
The training and validation database had a size of $20,000 \times n$ and $2,000 \times n$, respectively, while the test database had a size of $6,000 \times n$, where $n$ is the number of phasor measurements.
% This method is repeated to generate an input dataset of approximately 28000 samples. 

The performance of five ML models was investigated in this study. The first two models are LR and SVR, with the latter implemented
% models, namely, LR and SVR, are well-known in the power systems community; the latter was implemented 
using a radial basis function kernel. The third model was an Indirect DNN. This model first performs state estimation from sparsely-placed PMUs using the methodology proposed in \cite{azimian2022state}, and then computes the flows and injections from the outputs of the state estimator.
The fourth model was the Direct DNN described in Section \ref{DNN_Unobservability}, while the last model was the PIC-DNN developed in Section \ref{PIC-DNN}.
The hyperparameters used for the three DNN models are given in Table \ref{hyp_table}.
The loss function for the Direct and Indirect DNN models is the MSE, while for the PIC-DNN model we used \eqref{eq6}, with $n_{bin}  =  5$.
% The parameters are obtained by heuristic approaches. 
% We set the number of layers in the DNN models based on the direct correlation of the output features. 
Note that the number of neurons in each hidden layer is kept flexible (using the scaling factor, $\eta$) to account for the increase in number of input features due to increase in number of PMUs (see Section \ref{CS-2}).

% The input features are scaled with a scaling factor $\eta$ to get the number of neurons. Empirical tests show a value of $1 \leq \eta \leq 1.5$ to be optimal for the scaling factor. The other parameters are based on established models proposed in the literature.

\begin{table}[ht]
\caption{Hyperparameters of DNNs created in this study}
\label{hyp_table}
\centering
\begin{tabular}{|
>{\columncolor[HTML]{EFEFEF}}c |c|}
\hline
{\textbf{Parameter}} &
\cellcolor[HTML]{EFEFEF}{\textbf{Value}} \\ \hline

{\begin{tabular}[c]{@{}c@{}}Layers \\ \end{tabular}}  & 3  \\ \hline

{\begin{tabular}[c]{@{}c@{}}Neurons in  
each hidden layer\\ \end{tabular}} & \cellcolor[HTML]{FFFFFF} Number of input features $ \times \eta$ \\ \hline

{\begin{tabular}[c]{@{}c@{}}Learning rate\\ \end{tabular}}    & 1e-3  \\ \hline

{\begin{tabular}[c]{@{}c@{}} Number of  epochs\\ \end{tabular}}         & 200 \\ \hline

{\begin{tabular}[c]{@{}c@{}}Activation  function\\  \end{tabular}} & Rectified Linear Unit  \\ \hline

% {\begin{tabular}[c]{@{}c@{}}Loss metric \\ \end{tabular}} & Mean squared error (MSE)  \\ \hline

{\begin{tabular}[c]{@{}c@{}}Optimizer  \\  \end{tabular}} & Adam  \\ \hline

{\begin{tabular}[c]{@{}c@{}}Batch size  \\ \end{tabular}} & 64  \\ \hline

\end{tabular}

\end{table}

The simulations done to compare the performance of the five ML models were conducted on a computer with an Intel Core (TM) i7-11800H CPU @2.3GHz with 16GB of RAM and an RTX 3070Ti GPU.
The following two case-studies were designed.
% Two case studies are designed to assess the reliability and performance of the PIC-DNN model. 
The first case-study is a comparative analysis between PIC-DNN and other ML models when PMUs are placed on only the 11 HV buses of the test system. 
In the second case-study, the performance of all five ML models is investigated as more PMUs are added to the system. 

\vspace{-1em}
\subsection{Case-Study I: Performance Comparison with PMUs placed only on HV buses}
\label{CS-1}
In this case-study, the inputs to the ML models are the voltage and currents obtained from PMUs placed on the 11 HV buses of the IEEE 118-bus system.
Therefore, the number of phasor measurements is 41 (= 11 voltage phasors + 30 current phasors), i.e., $n=82$, since one phasor has two components, while the number of active power flows and injections are 490 (= 372 power flows + 118 power injections), i.e., $m=490$.
To ensure the consistency of estimates obtained from the five ML models, the analysis was repeated 100 times with different random subsets of the test dataset. 

The statistics of the error metric (RMSE) for the five ML models are recorded in Table \ref{cs1t1}. 
The results show that PIC-DNN does better than the other ML models in both the mean value of RMSE across all the flows and injections as well as the variation of RMSE over the 100 trials.
For example, the proposed PIC-DNN outperforms others in terms of the mean by at least $15\%$ and in terms of standard deviation by at least $40\%$; this denotes a significant improvement in both accuracy and consistency of power flow and injection estimates.
The poor performance of the Indirect DNN model can be attributed to the quadratic relationship between voltage and power, as highlighted in Section \ref{section2}. 
The relatively poor performance of SVR is due
% can be attributed 
to the input dataset being restricted to $5,000$ samples. This was done to compensate for its (longer) training time, indicating the handicap of using SVR for large datasets as its scalability is a concern \cite{abdiansah2015time}.

Lastly, note that the results shown in Table \ref{cs1t1} (as well as the last column of Table \ref{table1} and Fig. \ref{casestudy2p1}) were obtained when a two-component GMM noise was added to the voltage and current phasors. The mean, standard deviation, and weights of this noise for magnitudes and angles are $(-0.4\%, 0.6\%)$ and $(-0.2\degree, 0.3\degree)$, $(0.25\%, 0.25\%)$ and $(0.12\degree, 0.12\degree)$, and $(0.4, 0.6)$, respectively. These values corresponded to a $1\%$ TVE error for voltage and current phasors, as mandated in the IEEE/IEC Standard for PMUs \cite{PMU2018Standard}.

\begin{table}[hb]
% \vspace{-1em}
\caption{Performance comparison of ML models for power estimation in terms of their root mean square error (RMSE)}
\label{cs1t1}
\centering

\begin{tabular}{|
>{\columncolor[HTML]{EFEFEF}}c |c|c|}
\hline
{\textbf{ML Model}} & 
\cellcolor[HTML]{EFEFEF}{\textbf{Mean of RMSE}} & \cellcolor[HTML]{EFEFEF}{\textbf{\begin{tabular}[c]{@{}c@{}}Standard deviation of RMSE\\ \end{tabular}}} \\ \hline

{\begin{tabular}[c]{@{}c@{}}LR\\  \end{tabular}}  & 5.80 & 0.020 \\ \hline

{\begin{tabular}[c]{@{}c@{}}SVR\\  \end{tabular}} &  7.75 &  0.120 \\ \hline

{\begin{tabular}[c]{@{}c@{}}Indirect DNN\\ \end{tabular}}    & 8.37 & 0.020 \\ \hline

{\begin{tabular}[c]{@{}c@{}}Direct DNN \\ \end{tabular}}         & 5.91 & 0.024\\ \hline
{\begin{tabular}[c]{@{}c@{}}PIC-DNN \\\end{tabular}} & 4.92  & 0.012\\ \hline
\end{tabular}
\vspace{-1em}
\end{table}
% To confirm that the proposed PIC-DNN has higher consistency in power estimates, the errors in power flows and bus injections are compiled into a distribution plot for each model, given in \autoref{casestudy1p2}. The proposed PIC-DNN has a lower spread of estimation errors across buses than other models. The high variation of errors in indirect DNN is due to the problems discussed in \autoref{section2}. 
\vspace{-0.5em}
\subsection{Case-Study II: Impact of Increase in Number of PMUs}\label{CS-2}
% https://www.overleaf.com/project/6260eb74f03356a402332f02
% Analyses up until now have focused on placing PMUs on the high-voltage buses of the 118 bus network. 
In this case-study, the number of locations where PMUs must be placed is increased one at a time for each of the five ML models to determine how their performance changes with an increase in the number of PMUs in the IEEE 118-bus system.
This is a very practical scenario since, with additional investment in grid modernization, power utilities will add more PMUs not only in the lower voltages of the transmission system but also in the distribution system.
Therefore, this case-study indicates the improvements in ML-based power flow and injection estimation with an increase in PMU coverage.

\begin{figure}[ht]
%  \vspace{-1em}
	\centering
	\includegraphics[width=0.488\textwidth]{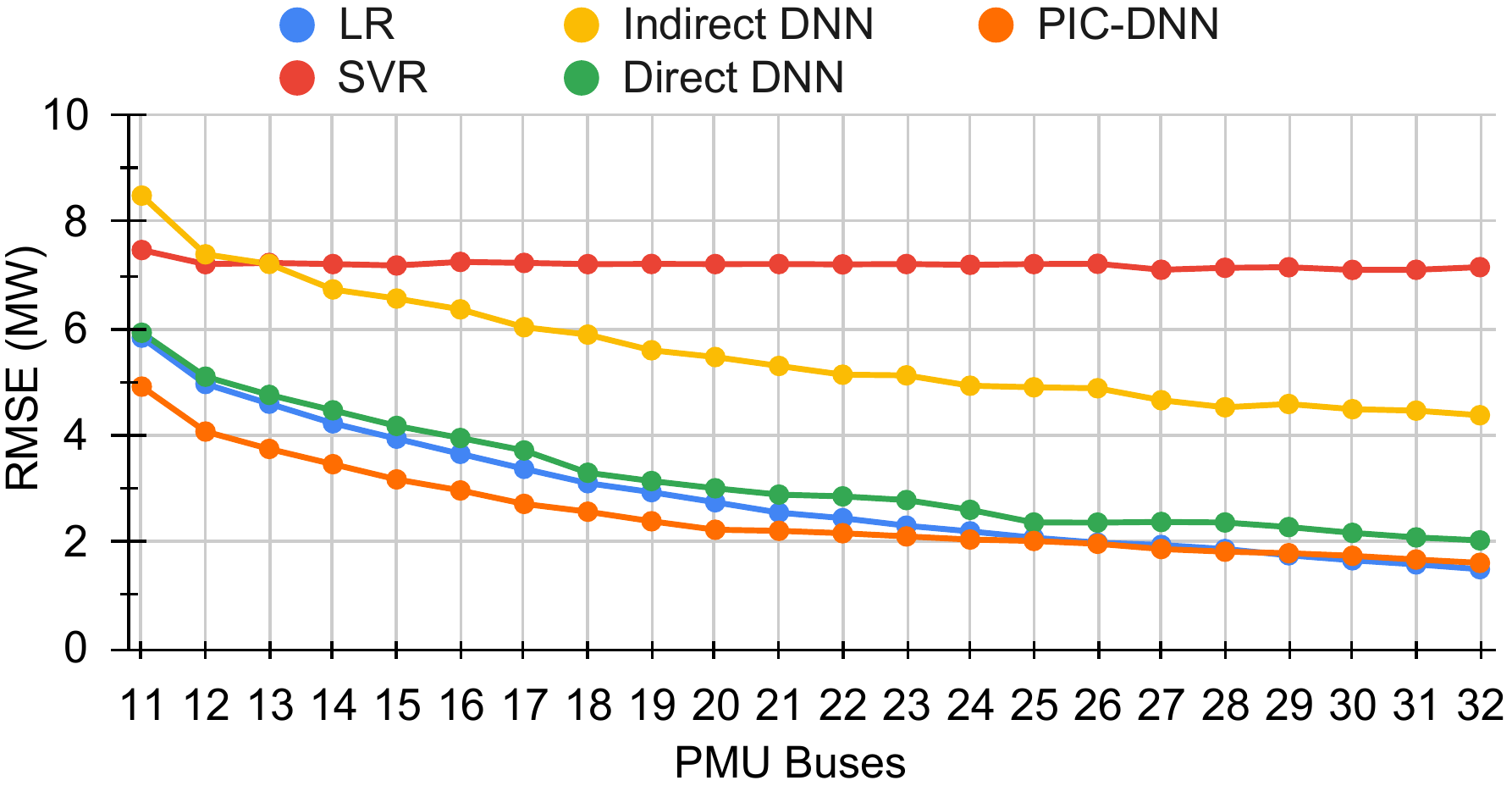}
 \vspace{-2em}
	\caption{Performance comparison of ML models with increase in number of PMUs}
	\label{casestudy2p1}
  \vspace{-1em}
\end{figure}

The results obtained as the number of locations where PMUs must be placed increased from 11 to 32 is shown in Fig. \ref{casestudy2p1}. 
To ensure fairness in comparison, the next optimal bus location for placing PMUs is determined independently for each ML model. This means that the horizontal axis of Fig. \ref{casestudy2p1} denotes the total number of buses where PMUs are placed; however, the bus names that go with that number can be different for the five ML models. 
We stopped at 32 because by placing PMUs on that many buses, we could directly compare the performance of the different ML models with the conventional LSE results (shown in Table \ref{table1}).

From Fig. \ref{casestudy2p1}, it can be realized that the proposed PIC-DNN performs better than the other four ML models for higher degrees of unobservability (fewer PMUs).
With an increase in the number of PMUs, the performance of LR becomes comparable with the proposed approach (and even better than PIC-DNN for PMUs placed at 30 or more bus locations). This is expected because linear models can successfully express the relations between the inputs and the outputs as the number of sensors increase.
% with the increased number of sensors.
Lastly, the superiority of data-driven approaches is realized from the fact that when the number of bus locations where PMUs are placed is 32, both LR and PIC-DNN outperform LSE results with two-component GMM measurement noise by $35\%$ and $28\%$, respectively.

\section{Conclusion}\label{section5}
This paper proposes a physics-inspired ML approach that uses DNNs to quickly and accurately estimate all the power flows and injections directly from PMUs placed in the transmission system. 
% The proposed PIC-DNN model ensures that the law of energy conservation is satisfied and performs reliable estimation when there are very few PMUs in the system. 
The proposed PIC-DNN model not only performs consistent estimation when there are very few PMUs in the system but also ensures that the law of conservation of energy is always satisfied.
The former is established by intelligently combining inferences drawn from historical SCADA data with real-time PMU data, while the latter is secured by adding a linear constraint to the loss function of the DNN. The results indicate that by placing PMUs on \textit{only 11 buses of the IEEE 118-bus system, PIC-DNN can estimate the flows and injections to within 5 MW of the actual value}, even in the presence of non-Gaussian noise in the PMU measurements. When the PMU locations increase to 32, the estimation error of the proposed approach becomes lower than 2 MW.
% The ability of PIC-DNN to handle non-Gaussian noise in PMU measurements is also demonstrated.
% The results indicate ...

% In this paper, we proposed a modified DNN approach that simultaneously estimates power flows and injections in a partially observable transmission system. The problem of current state estimation methods involves using slow timescale measurements that are not time synchronized. Furthermore, the scarcity of fast-time synchronized PMU measurements in power networks makes accurate estimation difficult. Further, the indirect estimation of power from power system states leads to a compounding of errors due to the quadratic relationship between voltage and power. This encouraged a direct method for power estimation using data-driven models, especially DNNs, which are known for their universal approximation capabilities and resiliency to non-Gaussian noise under severe unobservability. Leveraging the linear correlations between the power flows and injections, a \textit{physics inspired constrained DNN} is proposed that estimates the power flows and injections with the linear constraints implicitly included in the overall optimization problem. The proposed approach is better than other power estimation methods for partially observed networks. 

The focus of this paper has been on the transmission system. In the future, we plan to extend our analysis to distribution systems where PMUs are gradually being added and for which fast detection of reverse power flows is crucial. 
We will also combine the outcomes of this research to perform a better security assessment of the power system (e.g., by extending the research of \cite{biswas2021mitigation}).
The DNN models analyzed here had fully-connected feed-forward architectures. We are currently investigating the ability of graph neural networks to better incorporate the physical properties of the power system into ML training and execution (e.g., for ensuring robustness during topology changes).

\bibliography{bibtex.bib}

\begin{thebibliography}{10}
\providecommand{\url}[1]{#1}
\csname url@samestyle\endcsname
\providecommand{\newblock}{\relax}
\providecommand{\bibinfo}[2]{#2}
\providecommand{\BIBentrySTDinterwordspacing}{\spaceskip=0pt\relax}
\providecommand{\BIBentryALTinterwordstretchfactor}{4}
\providecommand{\BIBentryALTinterwordspacing}{\spaceskip=\fontdimen2\font plus
\BIBentryALTinterwordstretchfactor\fontdimen3\font minus
  \fontdimen4\font\relax}
\providecommand{\BIBforeignlanguage}[2]{{%
\expandafter\ifx\csname l@#1\endcsname\relax
\typeout{** WARNING: IEEEtran.bst: No hyphenation pattern has been}%
\typeout{** loaded for the language `#1'. Using the pattern for}%
\typeout{** the default language instead.}%
\else
\language=\csname l@#1\endcsname
\fi
#2}}
\providecommand{\BIBdecl}{\relax}
\BIBdecl

\bibitem{chen2020tracing}
Y.~C. Chen and S.~V. Dhople, ``Tracing power with circuit theory,'' \emph{IEEE
  Transactions on Smart Grid}, vol.~11, no.~1, pp. 138--147, 2020.

\bibitem{biswas2020graph}
R.~S. Biswas, A.~Pal, T.~Werho, and V.~Vittal, ``A graph theoretic approach to
  power system vulnerability identification,'' \emph{IEEE Transactions on Power
  Systems}, vol.~36, no.~2, pp. 923--935, 2020.

\bibitem{azmi2017evolution}
A.~Azmi, J.~Jasni, N.~Azis, and M.~A. Kadir, ``Evolution of transformer health
  index in the form of mathematical equation,'' \emph{Renewable and Sustainable
  Energy Reviews}, vol.~76, pp. 687--700, 2017.

\bibitem{amini2014determination}
M.~Amini, A.~I. Sarwat, S.~Iyengar, and I.~Guvenc, ``Determination of the
  minimum-variance unbiased estimator for {DC} power-flow estimation,'' in
  \emph{IECON 2014-40th Annual Conference of the IEEE Industrial Electronics
  Society}.\hskip 1em plus 0.5em minus 0.4em\relax IEEE, 2014, pp. 114--118.

\bibitem{al2016estimating}
A.~Al-Digs, S.~V. Dhople, and Y.~C. Chen, ``Estimating feasible nodal power
  injections in distribution networks,'' in \emph{2016 IEEE Power \& Energy
  Society Innovative Smart Grid Technologies Conference (ISGT)}.\hskip 1em plus
  0.5em minus 0.4em\relax IEEE, 2016, pp. 1--5.

\bibitem{deka2020joint}
D.~Deka, M.~Chertkov, and S.~Backhaus, ``Joint estimation of topology and
  injection statistics in distribution grids with missing nodes,'' \emph{IEEE
  Transactions on Control of Network Systems}, vol.~7, no.~3, pp. 1391--1403,
  2020.

\bibitem{chatterjee2015partitioned}
P.~Chatterjee, A.~Pal, J.~S. Thorp, and J.~De~La~Ree, ``Partitioned linear
  state estimation,'' in \emph{2015 IEEE Power \& Energy Society Innovative
  Smart Grid Technologies Conference (ISGT)}, 2015, pp. 1--5.

\bibitem{varghese2022transmission}
A.~C. Varghese, A.~Pal, and G.~Dasarathy, ``Transmission line parameter
  estimation under non-{G}aussian measurement noise,'' \emph{IEEE Transactions
  on Power Systems}, pp. 1--16, 2022.

\bibitem{wang2018assessing}
S.~Wang, J.~Zhao, Z.~Huang, and R.~Diao, ``Assessing {G}aussian assumption of
  {PMU} measurement error using field data,'' \emph{IEEE Transactions on Power
  Delivery}, vol.~33, no.~6, pp. 3233--3236, 2018.

\bibitem{pal2013pmu}
A.~Pal, G.~A. Sanchez-Ayala, V.~A. Centeno, and J.~S. Thorp, ``A {PMU}
  placement scheme ensuring real-time monitoring of critical buses of the
  network,'' \emph{IEEE Transactions on Power Delivery}, vol.~29, no.~2, pp.
  510--517, 2013.

\bibitem{pal2017PMU}
A.~Pal, A.~K.~S. Vullikanti, and S.~S. Ravi, ``A {PMU} placement scheme
  considering realistic costs and modern trends in relaying,'' \emph{IEEE
  Transactions on Power Systems}, vol.~32, no.~1, pp. 552--561, 2017.

\bibitem{anderson2012minimum}
J.~E. Anderson and A.~Chakrabortty, ``A minimum cover algorithm for {PMU}
  placement in power system networks under line observability constraints,'' in
  \emph{2012 IEEE Power and Energy Society General Meeting}, 2012, pp. 1--7.

\bibitem{khodayar2021deep}
M.~Khodayar, G.~Liu, J.~Wang, and M.~E. Khodayar, ``Deep learning in power
  systems research: A review,'' \emph{CSEE Journal of Power and Energy
  Systems}, vol.~7, no.~2, pp. 209--220, 2021.

\bibitem{abdiansah2015time}
A.~Abdiansah and R.~Wardoyo, ``Time complexity analysis of support vector
  machines ({SVM}) in {L}ib{SVM},'' \emph{International Journal Computer and
  Application}, vol. 128, no.~3, pp. 28--34, 2015.

\bibitem{azimian2021time}
B.~Azimian, R.~S. Biswas, A.~Pal, and L.~Tong, ``Time synchronized state
  estimation for incompletely observed distribution systems using deep learning
  considering realistic measurement noise,'' in \emph{2021 IEEE Power \& Energy
  Society General Meeting}.\hskip 1em plus 0.5em minus 0.4em\relax IEEE, 2021,
  pp. 1--5.

\bibitem{azimian2022state}
B.~Azimian, R.~S. Biswas, S.~Moshtagh, A.~Pal, L.~Tong, and G.~Dasarathy,
  ``State and topology estimation for unobservable distribution systems using
  deep neural networks,'' \emph{IEEE Transactions on Instrumentation and
  Measurement}, vol.~71, pp. 1--14, 2022.

\bibitem{zimmerman2010matpower}
R.~D. Zimmerman, C.~E. Murillo-S{\'a}nchez, and R.~J. Thomas, ``{MATPOWER}:
  Steady-state operations, planning, and analysis tools for power systems
  research and education,'' \emph{IEEE Transactions on Power Systems}, vol.~26,
  no.~1, pp. 12--19, 2010.

\bibitem{birchfield2017grid}
A.~B. Birchfield, T.~Xu, K.~M. Gegner, K.~S. Shetye, and T.~J. Overbye, ``Grid
  structural characteristics as validation criteria for synthetic networks,''
  \emph{IEEE Transactions on Power Systems}, vol.~32, no.~4, pp. 3258--3265,
  2017.

\bibitem{PMU2018Standard}
\emph{{IEEE/IEC} {I}nternational {S}tandard - {M}easuring relays and protection
  equipment - {P}art 118-1: {S}ynchrophasor for power systems -
  {M}easurements}, 2018.

\bibitem{biswas2021mitigation}
R.~S. Biswas, A.~Pal, T.~Werho, and V.~Vittal, ``Mitigation of saturated
  cut-sets during multiple outages to enhance power system security,''
  \emph{IEEE Transactions on Power Systems}, vol.~36, no.~6, pp. 5734--5745,
  2021.

\end{thebibliography}

\end{document}